\documentclass[twocolumn,showpacs,preprintnumbers,amsmath,amssymb,prl]{revtex4}

\usepackage{graphicx}
\usepackage{dcolumn}
\usepackage{bm}

\begin{document}

\newcommand{\half}{\mbox{$\frac{1}{2}$}}
\newcommand{\dyad}[1]{\mbox{\boldmath $#1$}}

\newcommand{\ddt}{\mbox{$\frac{d}{dt}$}}

\title{Quantum Mechanics and Reality are {\em Really} Local }

\author{Michael Clover}
\email{michael.r.clover@saic.com}
\affiliation{%
Science Applications International Corporation\\
San Diego, CA  }%

\date{\today}

\begin{abstract}
We attempt to pull together various lines of research whose ultimate conclusion points to the actual ``locality'' of Quantum Mechanics (QM).  We note that just as John Bell discovered various errors in previous ``proofs'' of the completeness of QM, he made an error of his own in deriving the ``non-locality'' of QM.  We show that QM satisfies the correct locality bound for non-commuting variables -- and is therefore local for 2 or more particles.  We further show that the entangled wavefunctions that produce non-local de-Broglie-Bohm guidance equations are an artifact of First Quantization, and that the wavefunctions describing such experiments do factorize in Second Quantization (QED).  
\end{abstract}

\pacs{03.65.-w, 03.65.Ud}
\maketitle

\section{\label{sec:background}Introduction}

Ever since the publication of John Bell's landmark papers on hidden variable theories and Quantum Mechanics~\cite{B66,B76} and his proof that local realistic hidden variable theories must satisfy an inequality  which Quantum Mechanics and experiment violate, it has been claimed that QM and reality are non-local.  Nonetheless, it has also been shown that this non-locality does not allow faster than light communication nor any other violations of Special Relativity~\cite{B75, E78}. Given this duck that does not waddle like a duck, nor quack like a duck, Occam's Razor suggests that maybe it's not a nonlocal duck.

\section{\label{sec:commutation}Another Bell Loophole}

Various authors~\cite{Sa91, GG02, Th96, mrc1} have proposed ``local'' models that exploit one of the known loopholes in Bell's Theorem -- either they assume unfair sampling or detector inefficiencies to discard certain events, or they require the 2 detectors of the experiment to be causally connected.  We now ask what loophole Quantum Mechanics uses, if it {\em is}  local, in order to evade Bell's inequality.

A few detailed examinations of the derivation of Bell's inequality~\cite{BMR92, R03, dBMR99}  have found that at some point, the derivation assumes that certain measurement operators commute.  Rizzi\cite{R03} has provided a paedogogical review of the error -- the implicit assumption of simultaneous measurements of a single particle's spin along different axes -- in the original Bell inequality;  De Baere {\it et al.}~\cite{dBMR99} have provided a more philosophical review of the assumptions of counter-factual reasoning that went into such proofs, as well as a demonstration that the process of constructing a Bell parameter relies on a particular operator identity that includes commutation relations. 

 If the Bell/CHSH\cite{CHSH69}  parameter is defined by the linear combination of experimental averages, 
\begin{eqnarray}
S_{Bell}^{expt} \equiv \langle {A}{B}\rangle  + \langle {A'}{B}\rangle + \langle {A}{B'}\rangle - \langle {A'}{B'}\rangle \ , \label{eq:bellparmx}
\end{eqnarray}
then quantum mechanics allows it to also be represented by the  expectation value of a multi-term measurement operator as well,
\begin{eqnarray}
S_{Bell}^{qm} \equiv \langle \psi | \hat{A}\hat{B}  +  \hat{A'}\hat{B} +  \hat{A}\hat{B'} -  \hat{A'}\hat{B'}| \psi \rangle \ . \label{eq:bellparmq}
\end{eqnarray}
If  Alice's $\hat{A}$ operators commute with Bob's  $\hat{B}$ operators  (the locality condition),  then the square of the Bell ``operator''  satisfies the following identity:
\begin{eqnarray}
\left( \hat{A}\hat{B}+\hat{A'}\hat{B} + \hat{A}\hat{B'} - \hat{A'}\hat{B'}\right)^2 
 & \equiv & 4\hat{I}  - [\hat{A},\hat{A'}] [\hat{B},\hat{B'}] \ , \nonumber
\end{eqnarray}
where  $\hat{A}^2=\hat{A'}^2 =\hat{B}^2 = \hat{B'}^2 = \hat{I}$,  given that the measurement operators have eigenvalues $\pm 1$.

For spin $\half$ particles, the measurement operators have the form
$\hat{A} = \hat{a}\cdot\vec{\sigma}$, 
where the $\vec{\sigma}$ are the Pauli spin matrices
 ($\sigma_i\sigma_j = i\epsilon_{ijk}\sigma_k \ ,  \ \sigma_i^2 = 1$)
  and the ${a_i}$ are the  direction cosines describing the Stern-Gerlach magnet orientation in the $z-y$ plane if the particles travel along the $\pm x$-axis.  In this case,  $[\hat{A},\hat{A'}] = 2i \sigma_x \sin(\alpha-\alpha') $. 
 Since both Stern-Gerlach magnets' axes are switched by $90^{\circ}$~\cite{FWL95, HLBBR03, BMR92}, the sines are unity  and the Bell parameter bound is given by
\begin{eqnarray}
|\hat{S}_{Bell}^2|  
 & \le  & \left| 4\hat{I} - (2 i \hat{\sigma_x})^2 \right|\ , \label{eq:fermion} \\
 \Rightarrow S_{Bell} & \le  & 2\sqrt{2}. \  \nonumber
\end{eqnarray}

For spin 1 photons, the operation of a polaroid filter or polarizing beam splitter aligned at an angle $\theta$ can be written as a projection operator:
\begin{eqnarray}
\dyad{P}(\theta) \hat{\epsilon} &=& \hat{\epsilon} \  , \ \ \ 
 \hat{\epsilon}                  = \left( \begin{array}{c} \cos\theta  \\
                                                                                                \sin\theta \end{array} \right) . \nonumber \\
\dyad{P}(\theta)  &=&\left( \begin{array}{cc} \cos^2\theta & \sin\theta \cos\theta  \\ 
                                                                       \sin\theta \cos\theta & \sin^2\theta    \end{array} \right)  \  , \nonumber \end{eqnarray}
We convert the projection operator with eigenvalues (0,1) into a measurement operator  with eigenvalues (-1,1) via,
\begin{eqnarray}
\dyad{A}(\theta)  &=&   2\dyad{P}(\theta)-1 =  \left( \begin{array}{cc} \cos2\theta & \sin2\theta   \\ 
                                                                        \sin2\theta  & -\cos 2\theta \end{array} \right) \   . \nonumber
\end{eqnarray}
$\dyad{A}$ can be recast into a form nearly identical to that for spin $\half$,
\begin{eqnarray}
\dyad{A}(\theta) = \cos 2\theta \left( \begin{array}{cc}
                                                                       1 & 0   \\ 
                                                                       0  &-1    \end{array} \right) 
                                                                        + \sin2\theta  \left( \begin{array}{ccc}
                                                                       0 & 1   \\ 
                                                                       1  &0    \end{array} \right) 
                                                                            +0   \left( \begin{array}{ccc}
                                                                       0 & i   \\ 
                                                                       -i  &0   \end{array} \right)   \ ,  \nonumber
\end{eqnarray}
resulting in a  Bell parameter for photons with the bound $|S_{Bell}^2| \le 4 - (2i\sigma_x)^2 \sin 2(\alpha-\alpha')\sin 2(\beta-\beta')$. Since photon polarizer angles are varied by multiples of $45^{\circ}$~\cite{OU88, WZ98}, the sines are again unity, and $S_{Bell} \le 2\sqrt{2}$.

\subsection{GHZ states}

The analog to singlet spin states of 2 fermions is the Greenberger-Horne-Zeilinger state of three or more particles, $\Psi = \frac{1}{\sqrt{2}}(|+ \ldots +\rangle  + i |- \ldots -\rangle)$, for which Mermin~\cite{M90} has defined a parameter similar to Bell's:
\begin{eqnarray}
F = \int d\lambda \rho(\lambda)\frac{1}{2i}\left[ \Pi_{j=1}^n (A_j + iA'_j) - \Pi_{j=1}^n (A_j - iA'_j)\right] \ . \label{eq:mermin}
\end{eqnarray}
Note that it is again implicitly assumed that both $A$ and $A'$ not only exist but can be measured at the same time for the same hidden variable.  The algebra for three particles yields the following identity for the square of the $F$ operator
\begin{eqnarray}
(\hat{F}_3)^2 &= 4 I & - [A_1,A_1'][A_2,A_2'] - [A_2,A_2'] [A_3,A_3'] \nonumber \\
                                      &  &-  [A_3,A_3'][A_1,A_1']  \nonumber \\
                             & = 4 I&  - 3 \, (2i\sigma_x \sin(a_j - a'_j))^2 \le 16 I\ , \nonumber \\
\Rightarrow  |F_3| &\le &4 \equiv 2^{3-1} \ , \nonumber
\end{eqnarray}
in agreement with Mermin's quantum result, $|F_n| \le 2^{n-1}$ and violating his ``locality'' bound, $|F_3^{loc}|\le 2$.  Three particles states are just as local and non-commutative as two particle states.

The implicit assumption of the  ability to simultaneously measure  spin eigenvalues along two different axes means that Bell's and Mermin's claim of  a  {\em locality}  bound is actually a  {\em commutativity} constraint.   It is  commutative hidden variable theories that are precluded by experiment.

\section{Entanglement and Non-Local Forces}

There is a second kind of non-locality that occurs in the de Broglie-Bohm (dBB) interpretation of multi-particle quantum mechanics, and which has been used to both  reject that interpretation and simultaneously confirm QM's non-locality (dBB is a {\em non-commutative} hidden variable theory).

 For a single particle, dBB assume that there is a hidden variable, position, for a particle moving with a non-hidden momentum.  Given a wavefunction, $\psi$, that is a solution to Schr\"odinger's equation, and given a (hidden) starting position $\vec{x}_0$, the particle follows a trajectory with a local velocity given by
\begin{eqnarray}
\vec{v} =  \frac{[\hat{x},\hat{H}]\psi(x,t)}{\psi(x,t)}  = - \frac{i\hbar}{m}\frac{\nabla \psi(x,t)}{\psi(x,t)}  = \vec{v}(x,t) \ , \label{eq:dbbtraj}
\end{eqnarray}
where the velocities (forces) are locally mediated by the wavefunction.  Schr\"odinger's wavefunction can instantaneously transmit information about the boundary conditions to the particle, but by using a relativistic wavefunction (Dirac's or Klein-Gordon's) one can include retardation effects and eliminate this type of non-locality.

In typical EPR experiments, the  two experimental particles are written in an ``entangled'' state,
\begin{eqnarray}
\Psi  =  \frac{1}{\sqrt{2}}(\psi(\vec{x}_1) \phi(\vec{x}_2) \pm \psi(\vec{x}_2) \phi(\vec{x}_1)) \ , \label{eq:tangled}
\end{eqnarray}
and the dBB prescription results in
\begin{eqnarray}
\vec{v}_1 = - \frac{i\hbar}{m}\frac{\nabla_1 \Psi(\vec{x}_1,\vec{x}_2)}{\Psi(\vec{x}_1,\vec{x}_2)}  = \vec{v}_1(\vec{x}_1,\vec{x}_2)  \ne \vec{v}_1(\vec{x}_1)\ , \nonumber
\end{eqnarray}
This dependence of one particle's velocity on another particle's instantaneous position {\em really} requires  spooky action at a distance, and persists even with relativistic wavefunctions. (Had the wavefunction factored, $\Psi = \psi(x_1)\phi(x_2)$, then each velocity would have been local function of its coordinates.)

Given that an entangled wavefunction  implies  non-local forces, the question we have to ask is whether such wavefunctions actually exist.  Or, just because we {\em can}  write an entangled wavefunction (equation~\ref{eq:tangled}),  do  we {\em have} to write it that way?   

We shall see that such entangled expressions are common in the context of first quantization, but in the context of second quantization (QED), wavefunctions actually take on product forms, and ``entanglement'' is  subsumed in the measurement operator instead of the wavefunction. 

\subsection{The Second Quantized Wavefunction}
 
 In one EPR experiment~\cite{OU88} that is atypical only in the level of theoretical analysis presented by the authors,
 Ou and Mandel consider the two-particle wavefunction in the context of QED,
\begin{eqnarray}
|\Psi\rangle &=& (T_x T_y)^{1/2}|x_1 y_2\rangle + (R_x R_y)^{1/2}|x_2y_1 \rangle \nonumber \\
            & & -i(R_y T_x)^{1/2}|x_1 y_1\rangle + i(R_xT_y)^{1/2}|x_2 y_2\rangle \  , \label{eq:omwf4} 
 \end{eqnarray}
with $T$ and $R$ denoting a beam-splitter's  transmission and reflection coefficients for the indicated polarization, $|x_1\rangle$ indicating an $x$-polarized photon heading toward detector 1,  etc.    They also use the polarized scalar field operators corresponding to (coefficients of) $e^{i\omega t}$,
 \begin{eqnarray}
 \hat{E}_1^{(+)}  &=&  \cos \theta_1  \hat{E}_s^{(+)} - i \sin\theta_1  \hat{E}_i^{(+)} \ , \label{eq:fieldamp} \\
 \hat{E}_2^{(+)} &=& i\cos \theta_2  \hat{E}_s^{(+)} + \sin\theta_2  \hat{E}_i^{(+)} \ , \nonumber \\
 \hat{E}_s^{(+)} & \propto & \vec{\epsilon}_x \hat{a}_s \ , \ \ \ \ \   \hat{E}_i^{(+)}  \propto \vec{\epsilon}_y \hat{a}_i \ , \nonumber 
\end{eqnarray}
with $\vec{\epsilon}_i$,  unit vectors pointing in the direction of the polarization and $\hat{a}_p$, photon destruction operators for signal or idler photons that satisfy $[\hat{a}_i , \hat{a}^{\dag}_j] = \delta_{ij}$. The probability of coincidence is related to the operator expectation value,
\begin{eqnarray}
P(\theta_1,\theta_2) = K \langle \Psi|  \hat{E}_1^{(-)} \hat{E}_2^{(-)} \hat{E}_2^{(+)} \hat{E}_1^{(+)} |\Psi\rangle \ , \nonumber
\end{eqnarray}
which evaluates to the quantum mechanical probability of detection,
\begin{eqnarray}
P(\theta_1,\theta_2) &=& K \left[ (T_x T_y)^{1/2} \cos\theta_1 \sin\theta_2  \right. \nonumber \\
                                       & &  + \left. (R_x R_y)^{1/2} \sin\theta_1 \cos\theta_2 \right] ^2 \ , \nonumber
\end{eqnarray}
and if $R_i = T_i = \frac{1}{\sqrt{2}}$, we have the typical expression $P\propto \sin^2(\theta_1 + \theta_2)$ while the expectation values that go into the Bell parameter take the form $\langle \hat{A}\hat{B}\rangle = \cos 2(\theta_1 + \theta_2)$.

Note that the four-term expression in equation~\ref{eq:omwf4} can be factored into
\begin{eqnarray}
|\Psi \rangle = \left[\sqrt{T_x} |x_1\rangle + i\sqrt{R_x} |x_2\rangle\right]    \left[\sqrt{T_y} |y_2\rangle - i\sqrt{R_y} |y_1\rangle\right] , \label{eq:prodwf}
\end{eqnarray}
a product form that will generate {\em local} dBB trajectories even though the (normalized) coincidence events violate the ``locality'' bound of Bell's inequality.

\subsection{Pre- and Post-Selection}

An experiment that   ``post-selects''  certain terms from a product-form wavefunction can allow local-realists to claim an efficiency loophole~\cite{dCG94}.  Kwait {\it et al.}~\cite{KZ95} devised an experimental method to use  type-2 parametric down-converters  to generate entangled photon beams directly, rather than using unentangled type-1 beams in a beam-splitter to generate the 4 combinations of Ou and Mandel.  Type-2 down-conversion creates a cone of signal rays (say $x$-polarized) and a cone of idler rays (say $y$-polarized); depending on the orientation of the crystal, the two emerging cones can be completely disjoint, or they can  intersect each other in one (tangential) or two points.  Assuming two intersection points (say 7 and 11 o'clock on the right-hand signal cone intersecting  5 and 1 o'clock respectively on the left-hand idler cone), it is impossible to say which photon is which -- first quantization would say the wavefunction describing those photons is entangled. Using only photons from  those intersection points (e.g. by aiming the fiber-optic leads at those points),  all the photons fed to an experiment are ``entangled'', {\it vs.} only half for Ou and Mandel.  

 From the 2nd quantization point of view, the wavefunction is still of a product form: $\Psi = \phi_x(\vec{k}_s)\phi_y(\vec{k}_i)\delta^3(\vec{k}_s + \vec{k}_i - \vec{K}_0)$, where the delta-function ensures momentum conservation with the incident photon ({\it e.g.} an idler photon emerging at 3 o'clock on its cone is paired with a signal photon at 9 o'clock on its cone, etc).  The  measurement operator  ``selects'' the particular emergent angles out of each cone of possibilities; if the fiber-optic leads were to be shifted to say 12 o'clock on the one cone and 6 o'clock on the other cone, the measurement operator would change but the wavefunction would remain the same. The ``post-selection'' of Ou and Mandel's method has merely been pushed earlier up the beam-line~\footnote{The efficiency loophole could also be made here, since true probabilities require normalization by all down-converted photons in the cone(s), not just those emitted at the intersections. Such probabilities would  satisfy Bell's original inequality~\cite{dCG94}. Our purpose  is not to show that non-localists have not plugged a loophole, but that the value of the limit they use is {\em wrong}.}.
 
 Even the ``entangled'' two electron decay from a singlet state that Bohm and Aharonov~\cite{BA57} analyzed should properly be written as a product in QED, since the wavefunction is a product of two isotropically emitted electrons: $\Psi = \psi_{+}(\vec{k}_1) \psi_{-}(\vec{k}_2)\delta(\vec{k}_1+\vec{k}_2)$; it is the detector's measurement operator that superposes the two possibilties analogous to equation~\ref{eq:fieldamp}.

\subsubsection{QED and Locality}

The ability to factor a multi-particle wavefunction into a product of functions of individual coordinates is required to make Bohmian  trajectories or guidance equations manifestly local. 
The analysis of EPR experiments in the context of  QED  shows that the wavefunctions do  factor into a product form with the ``entanglement'' contained in the measurement operators whose commutation rules determine what can be measured simultaneously.  QED is local.
  
 \subsubsection{QED and Completeness}
 
It is  in the context of QED that the de-Broglie-Bohm interpretation makes most sense:  the wavefunction describes a wave in an {\it aether}  of zero-point oscillators of $\half \hbar \omega$,  the particle is a relatively compact region of space where the amplitude of the wavefunction is $\sqrt{\frac{3}{2} \hbar \omega}$, and the centroid of that region of space follows the dBB guidance equations,
\begin{eqnarray}
\frac{d\vec{x}}{dt} & = & \frac{ \vec{j}(\psi(x))}{ \rho(\psi(x))} \ ,  \nonumber
\end{eqnarray}
where $\vec{ j}(\psi) = \frac{\hbar}{m}\Im ((\nabla\psi^{\dag}) \psi) - \psi^{\dag}(\nabla \psi ) $ and $\rho(\psi) = \psi^{\dag}\psi$.

The wave-particle duality of first-quantization is resolved into separate waves and particles in second-quantization at the same time the entangled wavefunction of first-quantization  becomes independent waves with entangled measurement operators in second-quantization.  In QED, one can think of the wavefunctions describing all the metaphysical possibilities open to the particle, while the pertinent combination of creation and destruction operators describes the epistemological possibilities of a given experiment.

With respect to the spin of a particle, dBB gives the guidance equation for the ``hidden'' components of spin analogously to the trajectory equation,~\ref{eq:dbbtraj}:
\begin{eqnarray}
\frac{ds_j}{dt} =\frac{ [s_j , H]\psi(x,t,\vec{s})}{\psi(x,t,\vec{s})} \propto  \epsilon_{jkl} B_k s_l \ , \nonumber
\end{eqnarray}
which means that as Alice's particle gets into the field of her Stern-Gerlach magnet, the two unmeasured components will start precessing, making it meaningless to ask what ``value'' they have while the third is being measured.  While  Alice and Bob can each measure/infer two components of spin, each pair of  components can only be considered to have been in a stationary state until one component was measured -- once Alice begins measuring her $y$ projection, the $z$ projection inferred from Bob's measurement is obsolete, and {\it vice versa}.
 
\section{Conclusion}

We have shown that the  Bell parameter is bounded by $2\sqrt{2}$ when the non-commutativity of the measurement operators is taken into account.  This means that the experiments that have claimed to show a non-local effect are only measuring a non-commutative effect and say nothing about the locality of quantum mechanics or reality.

We have also shown that when these EPR experiments are analyzed in the context of second quantization (QED), the wave-functions factor into a product form, allowing the de Broglie-Bohm guidance equations to be derived in a manifestly local manner for multiple as well as single particles.  With an equation of motion for all physical quantities -- whether they commute or not -- we have a complete interpretation of quantum mechanics. 

 To draw an analogy with another realm of physics, hidden variables are like quarks -- you can't see or measure them directly, but their existence explains so many other things that to deny them is irrational.

\bibliography{loc2}

\end{document}